\documentclass[11pt,twoside]{article}

\usepackage{asp2006}
\usepackage{epsf}
\usepackage{psfig}
\usepackage{lscape}

\usepackage{graphicx}  % SH

\begin{document}

\title{Dust Formation and Winds around Evolved Stars: \\
       The Good, the Bad and the Ugly Cases}   %%% Fill in title
\author{Susanne H{\"o}fner}   %%% Fill in author names
\affil{Department of Physics \& Astronomy,
       Division of Astronomy \& Space Physics,
       Uppsala University, Box 515, SE-75120 Uppsala, Sweden}    %%% Fill in author affiliations

\begin{abstract} %%% Abstract to run on from here.
Cool luminous giants, in particular asymptotic giant branch %(AGB)
stars, are among the most important sources of cosmic dust. Their
extended dynamical atmospheres are places where grains form and
initiate outflows driven by radiation pressure, leading to
considerable stellar mass loss and the enrichment of the
interstellar medium with newly-produced elements.
This review summarizes the current understanding of dust formation
and winds in such stars, sketching a system of criteria for
identifying crucial types of dust grains in the range of possible
condensates. Starting with an overview of the specific conditions
for dust formation in cool dynamic atmospheres, the role of grains
as wind drivers, as well as their influence on observable properties
of cool giants and the circumstellar environment is discussed in
some detail. Regarding the literature, special attention is given to
current developments, e.g., the debate concerning the Fe-content and
size of silicate grains in M-type AGB stars which are critical
issues for the wind mechanism, or recent advances in spatially
resolved observations and 3D modeling of giants and their dusty
envelopes.
\end{abstract}

%%% MAIN BODY OF TEXT GOES HERE. CONSULT "INSTRUCTIONS FOR AUTHORS USING
%%% LATEX2E MARKUP", SECTIONS 2.3-2.6 FOR HELP WITH EQUATIONS, FIGURES,
%%% AND TABLES.

%\section{}   %%% Top level section head (remove "%" symbol)
%\subsection{}   %%% Second level section head (remove "%" symbol)
%\subsubsection{}   %%% Lowest level section head (remove "%" symbol)
%\section*{}    %%% Unnumbered top level section head (remove "%" symbol)
%\subsection*{}   %%% Unnumbered second level section head (remove "%" symbol)

\section{Introduction}

Dust formation in evolved stars is a wide topic with many aspects,
and more than a few unknowns, despite several decades of research.
It is central to our understanding of the cosmic matter cycle, and,
consequently, connected to a range of other astrophysical questions,
e.g., the chemical evolution of galaxies, pre-solar dust grains, and
the formation of planetary systems. The atmospheres of cool evolved
giants and super-giants are sites where newly-produced chemical
elements can nucleate into dust grains which, in turn, may drive
massive outflows, enriching the interstellar matter with the
products of stellar nucleo-synthesis. Developments in observational
techniques such as space- and ground-based infra-red spectroscopy,
sub-millimeter observations and interferometry, combined with
laboratory work on grain materials, and progress in numerical
modeling have substantially improved the understanding of these
processes, but also pointed at gaps in our knowledge regarding
micro-physical and chemical aspects of dust formation.

Asymptotic giant branch (AGB) stars, i.e., low- and intermediate
mass stars in the final phases of their evolution, have to be
counted among the most important sources of cosmic dust, and in
particular carbon grains. Mass loss through radiation pressure on
dust particles eventually reaches such high rates that a
considerable fraction of the stellar mass is blown away, sealing the
fate of the star. With decreasing stellar mass, increasing
luminosity, and decreasing effective temperature, mass loss turns
into a runaway process where material is transported away from the
stellar surface faster than nuclear reactions change the stellar
interior. Finally, when the core of the star evolves into a white
dwarf, expelled material may briefly become visible as a planetary
nebula before it is dispersed into the interstellar
medium.\footnote{The occurrence of a planetary nebula phase depends
on how the timescale for the evolution of the stellar core compares
to the expansion of the circumstellar envelope. The lowest mass
stars probably evolve so slowly that the wind material is dispersed
before it can be ionized.}

Other types of evolved stars, i.e., red giant branch stars and cool
super-giants, also show signs of stellar winds. In these cases,
however, the role of dust and the actual driving mechanisms of their
outflows are more uncertain. In this review the discussion will
therefore be centered around AGB stars and structured into the
following main areas: First, the stage is set with an overview of
the specific conditions for grain formation and growth in the
complex environment of cool dynamic stellar atmospheres. Then the
role of the newly-formed dust grains as potential wind drivers is
discussed in some detail, followed by a brief overview of effects of
dust formation and mass loss on the circumstellar and interstellar
environment.

\section{Dust formation in cool giants: physical and chemical aspects} \label{s_basics}

While dust is an important component of many cool astrophysical
environments, some rather specific conditions apply to grain
formation in evolved stars. They can roughly be grouped into
chemical and dynamical aspects which, combined with prevailing
thermodynamic conditions, lead to characteristic properties of the
resulting dust particles.

Typical effective temperatures of AGB stars are around 3000~K, i.e.
quite low in stellar terms but not low enough to allow for
condensation of solids in the photospheric layers of the stars.
Therefore, dynamics is commonly assumed to play a key role in the
dust formation process. Shock waves -- caused by stellar pulsation
or large-scale convective motions -- propagate outwards through the
atmosphere and lift gas above the stellar surface, intermittently
creating dense, cool layers where solid particles may form. These
dust grains are then accelerated away from the star by radiation
pressure (provided that their radiative cross sections are high
enough), transmitting momentum to the gas through collisions, and
dragging it along. This chain of events supposedly leads to the
observed outflows with typical mass loss rates of $10^{-7} - 10^{-4}
\,$M$_{\odot}$/yr and wind velocities in the range of 5~--~30 km/s.

\subsection{Atmospheric composition and grain materials}

\begin{figure}[t]
\includegraphics
%[width=12cm] {hoefner_fig1.eps}
[width=12cm] {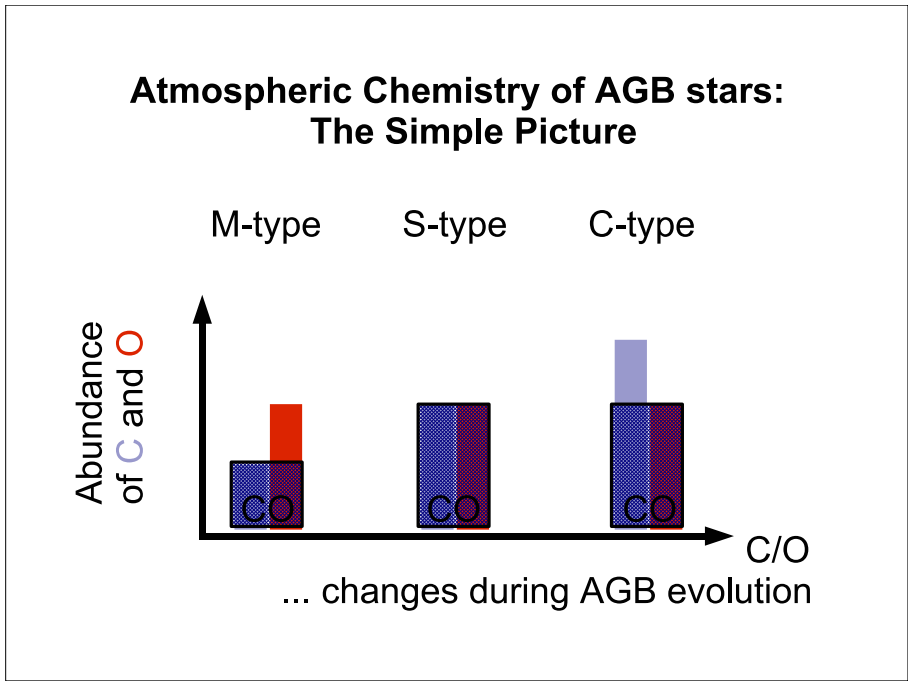}
\caption{Schematic picture of
atmospheric chemistry in AGB stars: the relative abundances of C and
O lead to three different cases due to the high bond energy of CO.
The less abundant element is completely bound in CO (cross-hatched
area), the excess O (M-stars) or C (C-stars) is available for other
molecules and dust grains.} \label{f_MSC}
\end{figure}

The raw material available for condensation of solids comes from two
sources, i.e., the original stellar matter -- product of previous
generations of stars -- which made up the star in question at the
time of its formation, and the products of nucleo-synthesis of the
individual star which are mixed to the surface by convection. As
both nucleo-synthesis and convection are ongoing processes, the
atmospheric composition, and, consequently, the dust composition
changes as the star evolves.

A crucial aspect in AGB stars is that the relative abundances of
carbon and oxygen change due to ongoing nuclear burning and
dredge-up of processed material, starting out with C/O $<$ 1 and
potentially reaching C/O $>$ 1 for stars within a certain initial
mass and metallicity range (cf. Lattanzio \& Wood 2004 for an
overview of stellar evolution on the AGB). The high bond energy of
the CO molecule leads to three different types of atmospheric
chemistries, corresponding to different spectral types (cf.
Fig.~\ref{f_MSC}):
\begin{itemize}
  \item C/O $<$ 1: all carbon is bound in CO, the excess oxygen is
  available for the formation of other molecules and dust (M-type);
  \item C/O $\approx$ 1: nearly all carbon and oxygen atoms are bound in
  CO, less abundant elements
  are crucial for the molecular composition of
  the gas, no abundant grain-forming materials are available
  (S-type);
  \item C/O $>$ 1: all oxygen is bound in CO, the excess carbon
  is mainly in the form of hydrocarbon molecules and carbon grains
  (C-type).
\end{itemize}

In order to predict which types of condensates will be dominant in
the different cases, some further aspects have to be taken into
consideration: Condensation temperatures of grain materials indicate
in which order different types of solids may form in response to
falling ambient temperatures as gas is moving away from the stellar
surface due to shocks or the onset of a wind. Gas densities and
element abundances regulate the nucleation and growth timescales of
grains, and, consequently the efficiency of condensation. According
to these criteria, the most abundant dust types should be those
which combine a high condensation temperature with high abundances
of their key ingredients. In practice, this means that silicates in
the form of olivine-type materials, i.e.
Mg$_{2x}$Fe$_{2(1-x)}$SiO$_4$ (with $0 \leq x \leq 1$), and/or
pyroxene-type materials, i.e. Mg$_{x}$Fe$_{(1-x)}$SiO$_3$, should be
dominant in M-type AGB stars, while C-stars should mainly produce
amorphous carbon grains and some SiC (see Gail 2003 for an in-depth
discussion of dust processes in AGB stars). These conclusions fit
well with observed features in infrared spectra of AGB stars.

\subsection{Dust formation: beyond the simple picture}

At this point, we need to add a few caveats which have implications
for the following discussion of dust grains as drivers or
(by-)products of stellar winds.

Regarding C-stars, it has to be noted that the term amorphous carbon
describes a whole class of materials with different microscopic
structures. Laboratory studies demonstrate that the ratio of sp$^2$
to sp$^3$ bonds in a specific sample will influence both the optical
properties and the bulk density of the grain material (see, e.g.,
J{\"a}ger et al. 1998). This, in turn, affects the dynamic
characteristics and observable properties of wind models (cf.
Andersen et al. 2003). A further source of uncertainty is the
relationship between the SiC component, inferred from the spectral
feature around 11 $\mu$m, and the featureless -- i.e.,
observationally not directly accessible -- amorphous carbon
component which, however, dominates the overall dust opacity and
determines the driving force of the wind. At present it is unclear,
whether these two types of condensates occur together or separately,
with a fixed ratio or not.

Concerning the composition of silicate grains in M-type AGB stars,
theoretical considerations favor olivine-type materials over
pyroxene-type condensates (e.g., Gail \& Sedlmayr 1999, Ferrarotti
\& Gail 2001) but observations are rather inconclusive on this
point. Furthermore, the Mg/Fe ratio of the grains has recently
become a matter of debate. From a simple gas kinetic point of view
(in contrast to chemical equilibrium condensation which is not
applicable in the dynamical atmospheres and winds of AGB stars, see
below) one would expect about equal amounts of Mg and Fe to be
included in the dust particles, corresponding to their roughly equal
abundances (for a solar mixture; cf. Gail \& Sedlmayr 1999). This
line of reasoning, however, ignores non-grey radiative effects on
grain temperature which will strongly favor the Fe-free end members
of the olivine and pyroxene sequences (cf. Woitke 2006b).

\begin{figure}[t]
\includegraphics
%[width=13cm] {hoefner_fig2.eps}
[width=13cm] {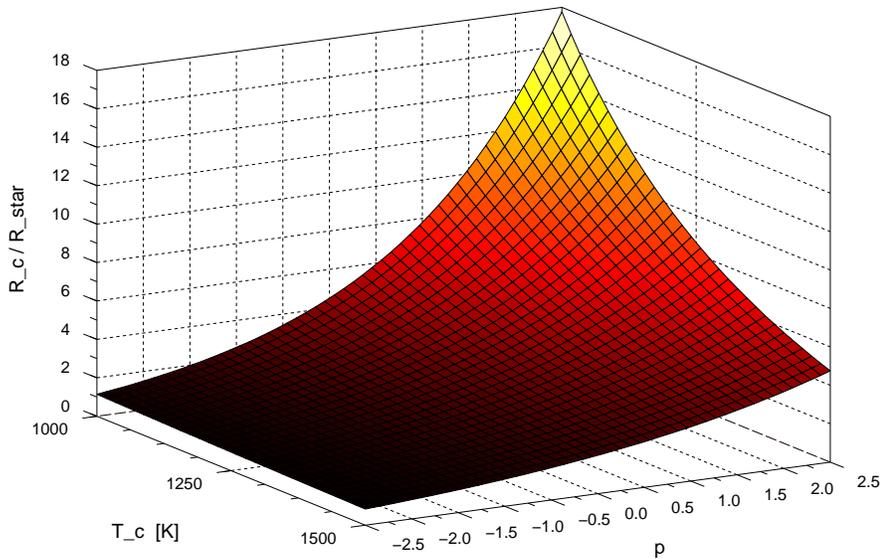}
\caption{Condensation distance $R_c$
(in units of the stellar radius $R_{\ast}$) as a function of the
power law index of the absorption coefficient $p$ around the stellar
flux maximum (see text) for a range of condensation temperatures
$T_c$ (assuming a stellar temperature of $T_{\ast} = 3000\,$K).
Silicates ($T_c \approx 1000\,$K) correspond to the lower end of the
temperature range plotted here, amorphous carbon particles ($T_c
\approx 1500\,$K) to the upper end. Representative values for $p$
are around 2 for olivine-type material with about equal amounts of
Fe and Mg, but close to $-1$ for Fe-free grains (forsterite),
resulting in $R_c > 10\,R_{\ast}$ and $R_c \approx 2-3\,R_{\ast}$,
respectively. Amorphous carbon corresponds to $p \approx 1$, leading
to condensation distances of $R_c \approx 2-3\,R_{\ast}$.}
\label{f_rc}
\end{figure}

The core of this issue can be understood with rather simple
analytical arguments: The growth and survival of dust particles
requires grain temperatures below the stability limit of the
respective condensate. The temperature of a grain will be mostly
determined by its interaction with the radiation field, and,
consequently, a condensation radius $R_c$ can be defined as the
distance from the star where the radiative equilibrium temperature
of a grain is equal to the condensation temperature $T_c$ of the
material, i.e. the closest distance where such grains can exist.
Assuming a Planckian radiation field, geometrically diluted with
distance from the star, and a power law for the grain absorption
coefficient $\kappa_{\rm abs}(\lambda)$ in the relevant wavelength
range (near $1\,\mu$m, around the flux maximum of the stellar
photosphere) the condensation radius can be expressed as
\begin{equation}\label{e_rc}
    \frac{R_c}{R_{\ast}} = \frac{1}{2} \left( \frac{T_c}{T_{\ast}} \right)^{- \frac{4+p}{2}}
    \qquad {\rm where} \qquad
    \kappa_{\rm abs} \propto \lambda^{-p}
\end{equation}
(for details see H{\"o}fner 2007 and references therein).
Figure~\ref{f_rc} shows the condensation distance $R_c$ (in units of
the stellar radius $R_{\ast}$) as a function of the power law index
of the absorption coefficient $p$ for a range of condensation
temperatures $T_c$ (assuming a stellar temperature of $T_{\ast} =
3000\,$K). For silicates, $T_c \approx 1000\,$K and $p$ is strongly
dependent on the Mg/Fe ratio. Considering olivine-type material,
grains with about equal amounts of Fe and Mg lead to $p \approx 2$
for the absorption coefficient of small grains and, consequently, to
$R_c/R_{\ast} > 10$, whereas iron-free particles with a
corresponding value of $p \approx -1$ can form at typically
$R_c/R_{\ast} \approx 2-3$ (comparable numbers are valid for
pyroxene-type particles). For amorphous carbon grains with $T_c
\approx 1500\,$K and $p \approx 1$ condensation distances are
comparable to Fe-free silicates, i.e., $R_c/R_{\ast} \approx 2-3$.
Despite the approximations used in these estimates, these numbers
compare well with detailed models.

The dependence of condensation distance on grain properties -- in
particular the strong variation of $R_c$ with Mg/Fe in silicates --
may have far-reaching consequences for the role of various dust
species as potential wind drivers or by-products of stellar outflows
as will be further discussed in the following sections. A few
general remarks, however, should be made already at this point.

A decisive but often ignored feature of dust condensation in stellar
atmospheres and winds is that low enough temperatures are a
necessary condition for the formation or survival of dust grains,
but not a sufficient one. Grain formation and growth rates are
dependent on prevailing gas densities and element abundances which
together affect the collision rates of condensible material in the
gas phase with other building blocks and existing grains. These
rates have to be compared to the typical dynamical timescales
connected with pulsations and outflows in order to determine which
types of dust particles will dominate in AGB stars. Dust materials
with low condensation rates caused by low abundances of key
ingredients or low condensation temperature (synonymous with low gas
density in the present context) are unlikely to form in the
time-critical environment of a cool giant's atmosphere or wind. As
matter moves away from the star, the temperature decreases which
favors condensation. At the same time, however, the nucleation and
growth of dust particles turns into a race against falling
densities. This results in non-equilibrium compositions of the gas
and dust phases and incomplete condensation (dust-to-gas ratios
different from chemical equilibrium values) in the outflow (see,
e.g., Ferraotti \& Gail 2001 for a more detailed discussion).
Consequently, the formation probability of grain materials decreases
drastically with increasing condensation radius, due to the strong
dependence of gas density (and therefore condensation rates) on
distance from the star.

Simple estimates based on gas kinetics as presented in Gustafsson \&
H{\"o}fner (2004) show that the timescales for the growth of carbon
grains close to the condensation radius are on the order of a year,
i.e. comparable with the pulsation period of the star, and
increasing outwards with falling density. This is in good agreement
with detailed numerical models which tend to show mean degrees of
condensation well below unity (typically 0.3-0.5), and a rather
limited zone of grain growth.

\section{Dust as a wind driver}\label{s_drivers}

\subsection{Basic principles}

In order to illustrate the interplay between atmospheric dynamics
and dust grains, the following simple model is adopted: The
dynamical evolution of a single mass element is described from an
instant when it has just been accelerated outwards by a passing
shock wave to the point where it is accelerated beyond the escape
velocity, or starts falling back towards the stellar surface.
Assuming that gravity and radiation pressure are the only relevant
forces (which is a reasonable approximation between shocks), the
equation of motion can be written as
\begin{equation}%\label{}
    \frac{du}{dt} =
       - g_0 \left( \frac{r_0}{r} \right)^2 \left( 1 - \Gamma \right)
    \qquad {\rm with} \qquad
    \Gamma = \frac{\kappa_H L_{\ast}}{4 \pi c G M_{\ast}}
\end{equation}
where $r$ is the distance from the stellar center, $u = dr/dt$ the
radial velocity, $g_0 = G M_{\ast} / r_0^2$ denotes the
gravitational acceleration at $r_0$, $\kappa_H$ the flux mean
opacity, and $M_{\ast}$ and $L_{\ast}$ are the stellar mass and
luminosity, respectively ($c =$ speed of light, $G =$ gravitation
constant). Atmospheric shocks due to stellar pulsation or
large-scale convective motions enter the picture by setting the
initial velocity $u_0$ of the matter element.

After the passage of a shock, the mass element first follows a
ballistic trajectory (since $\Gamma \ll 1$ in the hot dust-free gas
close to the photosphere), slowing down gradually due to gravity. If
no (or very little) dust is forming, $\Gamma$ will stay close to
zero and the matter will reach a maximum distance $r_{\rm max}$
according to its initial velocity,\footnote{In the discussion here
we assume that the velocity of the mass element is below the escape
velocity before the material reaches the dust condensation zone.
Otherwise it would be misleading to talk about a dust-driven wind.}
given by
\begin{equation} %\label{}
    \frac{r_0}{r_{\rm max}} = 1 - \left( \frac{u_0}{u_{\rm esc}} \right)^2
    \qquad {\rm where} \qquad
    u_{\rm esc} = \sqrt{\frac{2 G M_{\ast}}{r_0}}
\end{equation}
($u_{\rm esc} = $ escape velocity at $r_0$) and then fall back
towards the stellar surface.

\begin{figure}[tp]
\includegraphics
%[width=13cm] {hoefner_fig3a.eps}
[width=13cm] {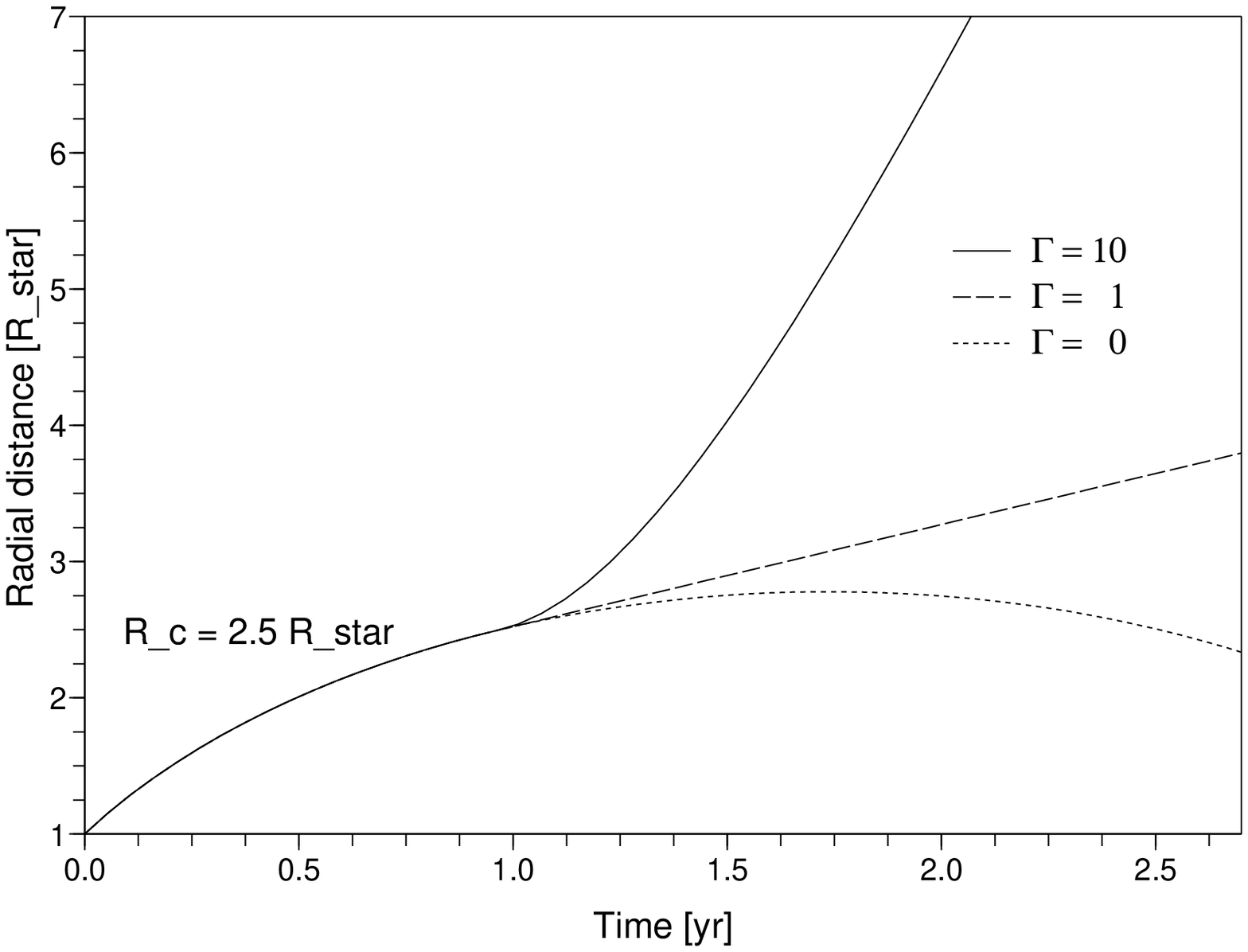}
\includegraphics
%[width=13cm] {hoefner_fig3b.eps}
[width=13cm] {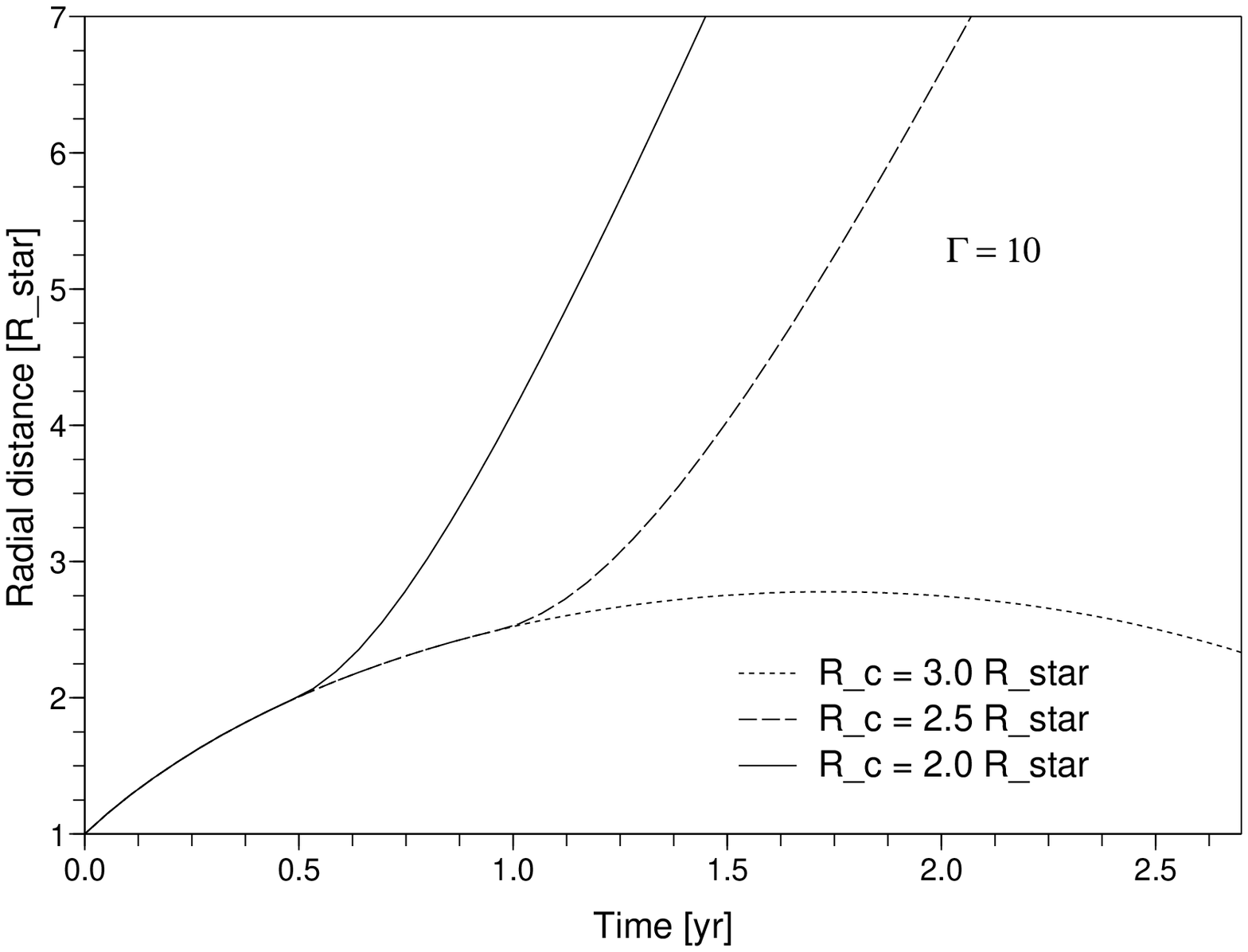}
\caption{A toy model for atmospheric dynamics, showing the location
of a matter element (i.e., its distance from the stellar center) as
a function of time. The velocity is determined by assuming a
distance-dependent $\Gamma$ (ratio of radiative acceleration to
gravitational acceleration) in the equation of motion, i.e., $\Gamma
= 0$ for $r < R_c$ (no dust), and $\Gamma =\,\,$constant at
distances beyond $R_c$. Both, the value of $\Gamma$ (varied in upper
panel, for constant $R_c$) and the condensation distance $R_c$
(varied in lower panel, for constant $\Gamma$) play a crucial rule
for the onset of a stellar wind. See text for details.}
\label{f_toy_gamma}
\end{figure}

As the formation and survival of dust grains requires temperatures
below the condensation temperature of the respective material,
grains will not exist closer to the stellar surface than their
corresponding condensation distance $R_c$
(cf.~Sect.~\ref{s_basics}). If the initial velocity imparted on the
gas element by the passing shock is high enough that 
the material reaches distances beyond $R_c$ 
grain formation and growth may set in, causing $\kappa_H$ and
thereby $\Gamma$ to increase. In this case, we can distinguish
several scenarios:
\begin{itemize}
\item
If $\Gamma > 1$, $du/dt$ is positive, the matter is accelerated away
from the star.
\item
If $\Gamma = 1$, the r.h.s of the equation of motion vanishes. The
matter element continues to move at a constant velocity.
\item
If $0 < \Gamma < 1$, the final fate of the matter element (escape or
fall-back) depends on the velocity at which it is moving when it
reaches the dust formation zone. As a positive value of $\Gamma$ can
be interpreted as a reduction of the effective gravitational
acceleration, the corresponding local escape velocity is smaller
than in the dust-free case, leading to a narrow regime where the
combined momentum transferred by the shocks and the radiation
pressure on dust drive the matter away from the star (cf.~H{\"o}fner
2007 for a more detailed discussion).
\end{itemize}
Figure~\ref{f_toy_gamma} illustrates the different scenarios (for
stellar parameters $M_{\ast} = 1\,M_{\odot}$, $L_{\ast} =
7000\,L_{\odot}$, $T_{\ast} = 2700\,$K, and initial conditions $r_0
= R_{\ast}$, $u_0 = 0.8 \, u_{\rm esc}$), assuming that $\Gamma = 0$
for $r < R_c$ (no dust), and $\Gamma =\,$constant at distances
beyond $R_c$. This is, of course, a simplification of the real
situation where dust formation and grain growth are spread out over
a range of distances (see H{\"o}fner 2008b for a more elaborate toy
model including grain growth), and, in addition, the resulting flux
mean opacity may vary due to other factors than the changing amount
of material condensed into grains.\footnote{Note that the optical
depth of the circumstellar envelope at wavelengths around the
stellar flux maximum may have a critical influence on the flux mean
opacity $\kappa_H$ and, consequently, on the variation of $\Gamma$
with distance from the star. As the optical depth approaches unity,
a matter element moving outwards will be increasingly shielded from
the direct stellar radiation, and will instead be exposed to the
flux emitted by the inner layers of the circumstellar envelope, with
the flux maximum shifting to larger wavelengths, corresponding to
lower temperatures. In a case where the dust opacity decreases with
increasing wavelength (e.g., for small amorphous carbon grains in
the near infrared) this will lead to a decrease of $\Gamma$ with
distance. As a high optical depth of the circumstellar envelope is
connected to high opacities and/or high densities, the shielding
effect will not be significant in winds close to the threshold for
dust-driven mass loss, but it may regulate the terminal velocity of
outflows corresponding to the highest mass loss rates. Also, it
should be noted that this is mostly relevant for C-rich AGB stars
(amorphous carbon grains), while envelopes of M-type objects usually
have comparatively low optical depths in the near infrared.} The
simple toy model, however, makes it easy to demonstrate that both,
the total opacity of the dust grains (i.e. the resulting $\Gamma)$
{\it and} the condensation distance $R_c$ play a crucial rule for
the onset of a stellar wind (see Fig.~\ref{f_toy_gamma}, upper and
lower panel, respectively). Even a small difference in $R_c$ for
different grain materials can have a dramatic effect on their
potential as a wind drivers.

\bigskip

While grain temperatures and, consequently, condensation distances
are determined by the respective absorption cross sections, the
opacities relevant for radiative pressure are a more complex topic.
The corresponding grain cross sections are given by a combination of
absorption and scattering, i.e.
\begin{equation}\label{e_cpr}
    C_{\rm rp} = C_{\rm ext} - g_{\rm sca} \, C_{\rm sca}
    = C_{\rm abs} + \left( 1 - g_{\rm sca} \right) \, C_{\rm sca}
\end{equation}
where $C_{\rm abs}$, $C_{\rm sca}$ and $C_{\rm ext}$ denote the
cross sections for absorption, scattering and extinction,
respectively, $g_{\rm sca}$ is the asymmetry factor describing
deviations from isotropic scattering ($g_{\rm sca} = 0$ for the
isotropic case, $g_{\rm sca}= 1$ corresponding to pure forward
scattering, i.e., no momentum being imparted on the grain due to
scattering; see, e.g., Kr{\"u}gel 2003), and $C_{\rm rp}$ may depend
severely on grain size, as well as on material properties.

In order to compute the total flux mean opacity $\kappa_H$,
determining the radiation pressure and, consequently, $\Gamma$, it
is necessary to know the combined opacity (cross section per mass)
of all grains of a certain size (i.e., radius $a_{\rm gr}$, for
spherical particles) in a matter element (with a total mass density
$\rho$),
\begin{equation}%\label{}
    \kappa_{\rm rp} (\lambda, a_{\rm gr}) =
    C_{\rm rp} (\lambda, a_{\rm gr}) \, n_{\rm gr}
    \,\, \frac{1}{\rho}
\end{equation}
($n_{\rm gr}$ = number density of grains of radius $a_{\rm gr}$) at
all relevant wavelengths $\lambda\,$. Introducing the efficiency
$Q_{\rm rp} = C_{\rm rp} / \pi a_{\rm gr}^2$ (radiative cross
section divided by geometrical cross section) which can be
calculated from refractive index data using Mie theory (cf., e.g.,
Bohren \& Huffman 1983) the dust opacity defined above can be
reformulated as
\begin{equation}\label{e_kpr_gen}
    \kappa_{\rm rp} (\lambda, a_{\rm gr}) =
    \pi \,
    a_{\rm gr}^2 \, Q_{\rm rp} (\lambda, a_{\rm gr}) \,\,
    n_{\rm gr}
    \, \frac{1}{\rho}
    =
    \frac{\pi}{\rho} \,\,
    \frac{Q_{\rm rp} (\lambda, a_{\rm gr}) }{a_{\rm gr}} \,\,
    a_{\rm gr}^3 \, n_{\rm gr}
\end{equation}
where $a_{\rm gr}^3 \, n_{\rm gr}$ represents the volume fraction of
a stellar matter element which is occupied by grains, apart from a
factor $4 \pi / 3$. This quantity can be rewritten in terms of the
space occupied by a monomer (basic building block) within the
condensed material, $V_{\rm mon}$, times the number density of
condensed monomers in the matter element under consideration,
$n_{\rm mon}$, i.e.,
\begin{equation}\label{e_vol_gr}
    a_{\rm gr}^3 \, n_{\rm gr}
         = \frac{3}{4 \pi} \, V_{\rm mon} \, n_{\rm mon}
         = \frac{3}{4 \pi } \, \frac{A_{\rm mon} m_p}{\rho_{\rm grain}} \,\,
            f_c \, \varepsilon_c \, n_{\rm H}
\end{equation}
where the monomer volume $V_{\rm mon}$ has been expressed in terms
of the atomic weight of the monomer $A_{\rm mon}$ and the density of
the grain material $\rho_{\rm grain}$ ($m_p$ = proton mass).
Assuming, for simplicity, that all grains at a given location have
the same size, the number density of condensed monomers $n_{\rm
mon}$ has been rewritten as the product of the abundance of the key
element of the condensate $\varepsilon_c$, the degree of
condensation of this key element $f_c$, and the total number density
of H atoms $n_{\rm H}$. Using $n_{\rm H} = \rho / (1 + 4 \,
\varepsilon_{\rm He}) \, m_p$ (where $\varepsilon_{\rm He}$ denotes
the abundance of He), we obtain
\begin{equation}\label{e_krp}
    \kappa_{\rm rp} (\lambda, a_{\rm gr}) =
       \frac{3}{4} \,\,\,
       \frac{A_{\rm mon}}{\rho_{\rm grain}} \,\,\,
       \frac{Q_{\rm rp} (\lambda, a_{\rm gr}) }{a_{\rm gr}} \,\,\,
       \frac{\varepsilon_c}{1 + 4 \, \varepsilon_{\rm He}} \,\,\,
       f_c \, .
\end{equation}
This way of writing the dust opacity which may seem unusual at first
has the advantage of distinguishing between material properties
(constants and optical properties) and the chemical composition of
the atmosphere/wind, and allows to take into account that $Q_{\rm
rp} (\lambda, a_{\rm gr}) / a_{\rm gr}$ becomes independent of grain
size for particles much smaller than the wavelength $\lambda$. For a
given grain material and stellar element abundances, the only
unknown quantity in this expression of the opacity is the degree of
condensation $f_c$ (i.e. the fraction of available material actually
condensed into grains). As dust formation in AGB stars usually
proceeds far from equilibrium, $f_c$, in general, needs to be
calculated with detailed non-equilibrium methods such as those
described by, e.g., Gail \& Sedlmayr (1988, 1999). Assuming a
reasonable value for $f_c$, however, this equation allows to
estimate which grain types can, in principle, drive an outflow, and
which alternatives can be ruled out for this purpose based on too
small radiative pressure.

Taking amorphous carbon as an example ($A_{\rm mon} = 12$), the
optical properties according to Rouleau \& Martin (1991) give
$Q_{\rm rp} (\lambda, a_{\rm gr}) / a_{\rm gr} \approx 2 \cdot
10^{4}\,\,$[1/cm] at $\lambda \approx 1 \mu$m (near the stellar flux
maximum) in the small particle limit, with a corresponding density
of the grain material $\rho_{\rm grain} = 1.85\,\,$[g/cm$^3$].
Assuming C/O = 1.5 at solar metallicity, leading to $\varepsilon_c =
3.3 \cdot 10^{-4}$ (abundance of excess carbon not bound in CO),
together with complete condensation of all available carbon ($f_c =
1$), $L_{\ast} = 5000 \, L_{\odot}$ and $M_{\ast} = 1 \, M_{\odot}$,
results in $\Gamma \approx 10$. This is clearly more than enough for
driving the wind of a typical AGB star, and even a small fraction of
the available carbon being condensed into grains (about 10-20
percent in this case) will be sufficient to cause an outflow.

\smallskip

\subsection{Detailed dust-driven wind models and observations}

Radiation pressure on dust grains has long been suspected to be the
driving force behind the slow, massive outflows observed in cool
giants, in particular for carbon stars (see, e.g., Dorschner 2003 or
Habing and Olofsson 2004 for historical overviews), but early
modeling attempts had to rely on crude assumptions regarding dust
properties and dynamics (not to mention radiative transfer and
molecular opacities) due to a lack of input data and insufficient
computational resources. Time-dependent dynamical models of
pulsating atmospheres and dust-driven winds where pioneered by Wood
(1979) and Bowen (1988), taking the effects of atmospheric shock
waves due to pulsation into account but reducing the description of
dust to simple parameterized opacities, without a treatment of grain
formation and destruction. Consequently, these models could only
predict mass loss rates and other wind characteristics but they
allowed no conclusions about grain types and dust yields.

Progress in computational resources and new data from laboratory
studies of dust materials eventually made it possible to adopt a
more detailed, time-dependent description of the dust component,
including nucleation, growth and destruction of grains, in dynamical
models of C-type AGB stars (using methods developed by, e.g., Gail
\& Sedlmayr 1988 and Gauger et al. 1990). At first, the application
of this improved dust treatment was restricted to stationary winds
(e.g. Dominik et al.~1990). Soon after, however, models combining
grain formation, growth and evaporation of particles with
time-dependent dynamics were developed (e.g. Fleischer et al.~1991),
which allowed to study the interaction of pulsation-induced shock
waves and dust. As mentioned above, dust processes depend critically
on ambient temperatures and densities. This sensitivity to
thermodynamical conditions in combination with the feedback of the
dust component into gas dynamics and the radiation field may lead to
the formation of discrete dust shells (e.g. Fleischer et al.~1992),
dynamical instabilities (e.g. Fleischer et al.~1995, H{\"o}fner et
al.~1995), multi- or non-periodic behavior of the circumstellar
envelope (e.g. Winters et al.~1994, H{\"o}fner \& Dorfi 1997), and
incomplete condensation (low dust-to-gas ratios). Another topic
investigated with this type of models is the role of grain drift
relative to the gas which may modify grain growth rates and the
efficiency of wind acceleration (e.g. Sandin \& H{\"o}fner 2004), as
well as causing flow instabilities which may explain large-scale
patterns in circumstellar envelopes (Simis et al.~2001) as observed
around IRC+10216 (Mauron \& Huggins 1999, 2000).

Mass loss rates predicted by such wind models have been used in
theoretical studies of AGB evolution (e.g. Schr{\"o}der et al.~1999,
2003, Wachter et al.~2002, 2008), and synthetic IR colors have been
applied to the interpretation of photometric observations
characterizing stellar variability and dust-driven mass loss (e.g.
Winters et al.~1997, 2000, Le Bertre \& Winters 1998). A closer
comparison of model-based spectra and observations, however,
revealed some serious shortcomings of this first generation of dusty
dynamical models. The crude, grey treatment of radiative transfer
and opacities in the dynamical computations proved to be a major
drawback, resulting in unrealistic atmospheric structures, and
thereby influencing the conditions for dust formation and the
observable properties predicted by the models.

As a consequence, a new generation of dynamic atmosphere and wind
models was developed, featuring a frequency-dependent treatment of
radiative transfer which includes detailed molecular and grain
opacities. A well-tested example are the models for C-rich AGB stars
by H{\"o}fner et al.~(2003). They compare favorably with such
diverse observable properties as low-resolution spectra covering
about two orders of magnitude in wavelength, with features formed at
different depths in the atmosphere and wind (Gautschy-Loidl et
al.~2004), and high-resolution spectra showing variations of
molecular line profiles which probe the dynamics of atmospheric
shocks and wind acceleration (Nowotny et al.~2005ab). Mattsson et
al.~(2007, 2008) have applied these models to investigate the
formation of detached shells in connection with a He-shell flash,
and the dependence of mass loss on metallicity.

\begin{figure}
\begin{center}
\includegraphics[angle=0, width=13cm]{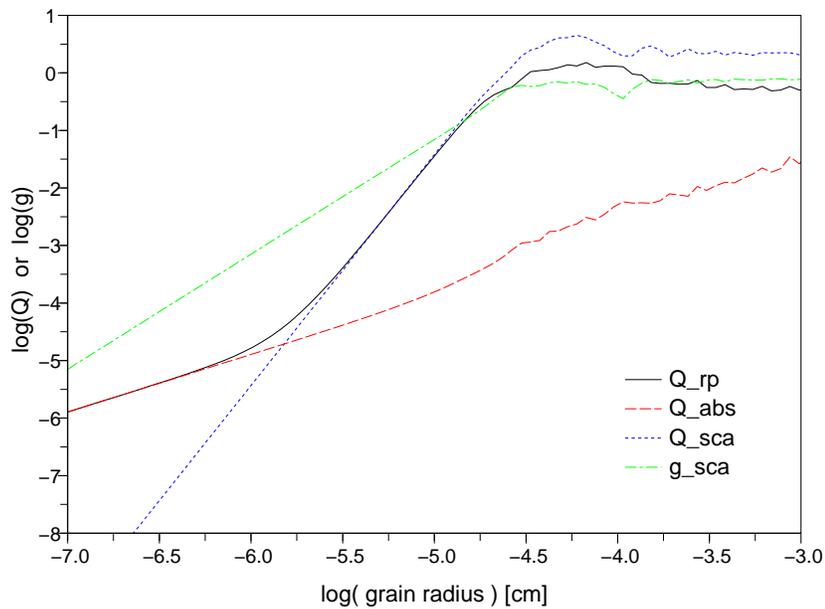}
\includegraphics[angle=0, width=13cm]{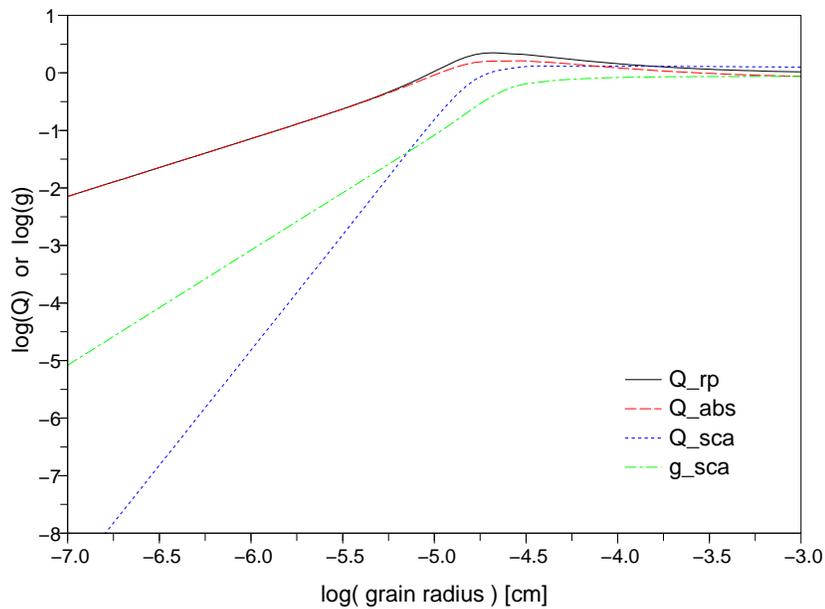}
\caption{ Optical properties of dust particles at wavelength
$\lambda = 1\,\mu$m as a function of grain radius: log($Q_{\rm rp}$)
(solid lines), log($Q_{\rm abs}$) (dashed), log($Q_{\rm sca}$)
(dotted) and log($g_{\rm sca}$) (dash-dotted), based on Mie theory
for spherical grains. Upper panel: forsterite particles
(Mg$_2$SiO$_4$). Lower panel: amorphous carbon grains. See text for
details.}\label{f_q_mie}
\end{center}
\end{figure}

The development of similar detailed models for winds of M-type AGB
stars has proved more difficult than in the C-rich case, not the
least due to a more complex chemistry of dust formation. Recently,
new models by Woitke (2006b) demonstrated that silicate grains
condensing in M-type atmospheres have to be basically Fe-free, since
even a low Fe-content will lead to much higher grain temperatures,
and, consequently, large condensation distances (see discussion in
Sect.~\ref{s_basics}). Without the inclusion of Fe in olivine- or
pyroxene-type grains, however, radiative pressures are too low for
driving a wind, as long as the grains are small compared to the
relevant wavelengths. For clarity, it has to be mentioned in this
context, that this problem was not apparent in earlier models with
grey radiative transfer, e.g., by Jeong et al.~(2003). Such grey
models will severely underestimate the temperature of silicate
particles which include Fe and give condensation distances
comparable to C-rich models, resulting in marginally sufficient
radiative driving (note that grey models correspond to $p = 0$ in
Fig.~\ref{f_rc}).

As a possible solution of this problem, H{\"o}fner (2008a)
investigated the viability of driving winds with micron-sized
Fe-free olivine-type grains (forsterite), instead of small
particles. For wavelengths corresponding to the flux maximum of AGB
stars, the radiative pressure experienced by a given total amount of
Fe-free olivine-type material is strongly dependent on particle
size, potentially exceeding the gravitational force for grain radii
just below $1 \mu$m at typical stellar parameters (cf. Fig.~1 in
H{\"o}fner 2008a). The strong variation with grain size is a
consequence of the shifting relative contributions of absorption and
scattering to the total cross section for radiative pressure (cf.
Eq.~\ref{e_cpr}). For particles which are small compared to the
wavelengths under consideration, scattering is negligible and this
cross section will be dominated by absorption which is very low for
Fe-free olivine-/pyroxene-type materials around wavelengths $\lambda
\approx 1\,\mu$m. The relative contribution of scattering, compared
to absorption, however, increases steeply with particle radius, and
starts to dominate the radiative pressure for this dust type long
before the grain radius approaches the wavelength, possibly bringing
up the cross section for radiative pressure to values required for
driving an outflow.

The grain size dependence of optical properties is illustrated by
Fig.~\ref{f_q_mie}, showing the efficiency factors for absorption,
scattering and radiative pressure, $Q_{\rm abs}$, $Q_{\rm sca}$ and
$Q_{\rm rp}$, respectively, as a function of grain radius
(cf.~Sect.~\ref{s_drivers} for definitions). The quantities are
calculated using Mie theory for spherical particles (program {\tt
BHMIE} from Bohren \& Huffman 1983, modified by B.T.~Draine).
%, {\tt www.astro.princeton.edu/\~{ }draine/scattering.html}).
Note that the behavior displayed by the silicate material (upper
panel; Mg$_2$SiO$_4$, refractive index data taken from J{\"a}ger et
al.~2003) differs crucially from the case of amorphous carbon (lower
panel; data of J{\"a}ger et al. 1998, sample cel1000) where
absorption dominates radiative pressure for particles of sizes up to
about 1 $\mu$m (due to the much higher level of $Q_{\rm abs}$), and
the dependence of $Q_{\rm rp}$ on grain size is much less
pronounced.

\begin{figure}[t]
%\centering {\includegraphics[angle=0, width=14cm]{hoefner_fig5.eps}}
\centering {\includegraphics[angle=0, width=14cm]{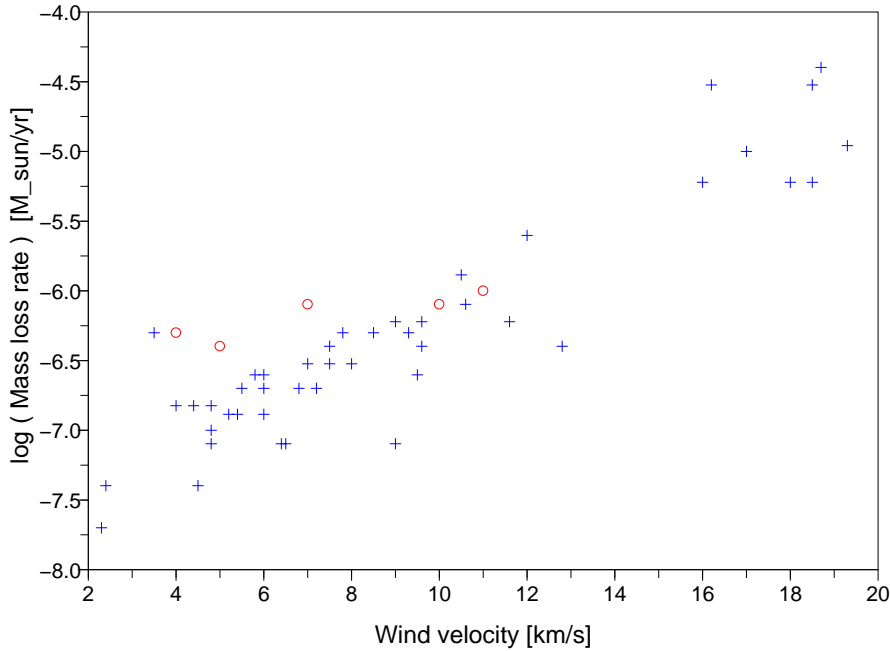}}
\caption{Mass loss rate vs. wind velocity for M-type AGB stars:
crosses mark observations by Olofsson et al. (2002) and Gonz{\'a}lez
Delgado et al. (2003), circles correspond to detailed RHD models by
H{\"o}fner (2008a) where the winds are driven by micron-sized
Fe-free silicate grains.}
   \label{f_dMdt_v}
\end{figure}

The detailed frequency-dependent atmosphere and wind models of
H{\"o}fner (2008a) which include a time-dependent treatment of
forsterite grains demonstrate that radiation pressure on such
particles is sufficient to drive outflows if prevailing conditions
allow grains to grow to sizes in the micro-meter range. The
resulting combinations of mass loss rates and wind velocities
compare rather well with observations of M-type AGB stars (cf.
Fig.~\ref{f_dMdt_v}) and preliminary test of synthetic spectra show
reasonably good agreement with observed near- and mid-IR spectra
(Lederer et al., in prep.).

\section{Dust as a (by-) product: features, patterns and yields}

In the previous section, the discussion has been focused on the
dominant dust species and their potentials as wind drivers. Now we
will take a look at other aspects of dust in evolved stars, i.e.,
the influence of grains on observable properties of these objects,
and estimated contributions from cool giants to interstellar dust.

The progress of space-based infrared spectroscopy achieved in recent
years, in particular at mid- and far-IR wavelengths, has had a
dramatic influence on our understanding of cool stellar winds.
Simultaneous coverage of a wide spectral range with the instruments
aboard the {\it Infrared Space Observatory} (ISO) has allowed an
analysis of molecular features originating
in different layers of 
stellar atmospheres and outflows, an unprecedented tool for studying
their structure and dynamics. A large variety of dust features has
been discovered in the infrared spectra of evolved stars, some
expected, some surprising (see, e.g., Dorschner 2003 or Molster \&
Waters 2003 for an overview). Examples include the well-studied SiC
feature at 11$\,\mu$m in C-type AGB stars, and the characteristic
features of silicates at about 10$\,\mu$m and 18$\,\mu$m (Si-O
stretching and O-Si-O bending modes), as well as more controversial
cases like the 13$\,\mu$m feature in M-type AGB stars which was
first attributed to corundum (Al$_2$O$_3$; Vardya et al.~1986, Onaka
et al.~1989, Begemann et al.~1997) but later more convincingly
identified with spinel (MgAl$_2$O$_4$; Posch et al.~1999, Fabian et
al.~2001). Some of the species discovered or predicted theoretically
are too rare or have too low opacity to contribute as wind drivers.
But certain high-temperature condensates, even if they are rare due
to low abundances of their key ingredients, may play an important
role as seed nuclei for olivine/pyroxene condensation in M-type AGB
stars (see, e.g., Gail \& Sedlmayr 1999, Jeong et al. 2003, Nuth \&
Ferguson 2006). Another interesting aspect is the discovery of
Mg-rich crystalline silicates (Molster et al.~2002abc) in addition
to their more abundant amorphous counterparts, since a highly
ordered microscopic structure of the condensate hints at a formation
at high temperature, close to the condensation limit.

In addition to spectroscopy, new ways of studying circumstellar dust
shells have become available through interferometry and imaging.
Present-day interferometers achieve resolutions down to the scales
of stellar radii, allowing detailed investigations of the innermost
parts of circumstellar envelopes and wind acceleration regions.
Certain features seen in near-IR spatial intensity distributions
have been interpreted as dense molecular layers in the upper
atmospheres and wind acceleration regions, probably related to shock
waves, based on their variations with wavelengths and phase (e.g.,
Tej et al.~2003, Weiner 2004, Woodruff et al. 2008, Wittkowski et
al. 2008). The inner edges of circumstellar dust shells seem to be
located at a few stellar radii (e.g., Danchi et al. 1994, Tevousjan
et al. 2004) while SiO masers are observed at somewhat closer
distances (e.g., Boboltz \& Wittkowski 2005, Wittkowski et al.
2007), in agreement with theoretical models. All interpretations of
such data in terms of radii/distances (with or without limb
darkening) should however be regarded with caution since many
observations indicate non-spherical stellar shapes or brightness
variations across the stellar surface (possibly due to large-scale
convective motions), as well as complex structures in the
circumstellar envelopes close to the star (e.g., Weigelt et al.
1998, 2002, Hofmann et al. 2000, Monnier et al. 2004, Ragland et al.
2008, Tatebe et al. 2008).

Such spatially resolved observations have stimulated the development
of 2D/3D dynamical models for AGB stars and their circumstellar
environments. The computational effort behind such models is
considerable, and, consequently, examples are a lot less numerous
than the spherically symmetric atmosphere and wind models discussed
above. There is a severe trade-off between generalized geometry and
physical/chemical approximations necessary in the models, as well as
serious limitations regarding the spatial ranges and time spans
covered by the 2D/3D simulations. Nevertheless, investigations of
the effects of intrinsically three-dimensional phenomena, like
convection or flow instabilities, on dust formation and mass loss
seem necessary in view of interferometric results, and the
simulations performed so far have provided valuable insights,
despite their limitations. Woitke (2006a) constructed 2D
(axisymmetric) dust-driven wind models with time-dependent dust
formation and grey radiative transfer. With these models, he
investigated how intrinsic instabilities in the dust formation
process create complex patterns in the circumstellar envelope
(without taking the pulsation of the central star into
consideration). Freytag \& H{\"o}fner (2008) on the other hand,
studied the effects of large-scale convective motions and of the
resulting shock waves in the atmosphere on time-dependent dust
formation in the framework of 3D %RHD
radiation-hydrodynamical 'star-in-a-box' models. The atmospheric
patterns resulting from convection are found to be reflected in the
circumstellar dust distribution, due to the strong sensitivity of
grain formation to temperatures and gas densities.

Looking, quite literally, at the dust and gas flowing from the star
into the surrounding environment, the question of yields comes to
mind. Compared to the task of determining the total amount of
newly-produced elements fed into the interstellar medium by AGB
stars, the problem of computing reliable dust yields has an extra
layer of complication. It is not sufficient to follow the evolution
-- in terms of nucleo-synthesis and stellar parameters -- of a
likely population of stars, including a realistic description of
mass loss. In addition, the fractions of elements condensing into
various types of dust grains have to be calculated along the
evolution tracks using a detailed time-dependent method since dust
formation proceeds far from equilibrium. The most comprehensive
modeling in this area so far has been done by Ferrarotti and Gail
(2006). They combined synthetic stellar evolution models with a
non-equilibrium dust formation description for the most abundant
grain materials in M-, S- and C-type AGB stars. Their dusty outflow
models, however, suffer from two major drawbacks. The mass loss rate
is an input parameter of their dynamical models, not a result, and
the flows are assumed to be stationary, neglecting the effects of
shock waves. A follow-up study with more realistic dynamical wind
models is far from trivial but urgently needed to provide reliable
input for models of the interstellar medium.

\section{Summary and conclusions}

If one would try to characterize the topic of dust formation in
evolved stars with one word the choice would probably fall on
`complex'. At a first glance it seems that there is an almost
infinite spectrum of possibilities to build solid particles from the
chemical elements available in cool giants, and have them interact
with dynamical and radiative processes in stellar atmospheres and
winds.

Fortunately, there is a system of criteria which helps to sort
materials, processes and interactions according to relevance for the
problem at hand, in particular when it comes to investigating wind
mechanisms. Abundances of chemical elements determine, in principle,
which types of condensates 
should be expected at a given stage of stellar evolution. Dynamical
processes like convection, pulsation, shocks, and outflows, on the
other hand, will often determine, in practice, which of the possible
grain types will actually form, as dynamics is setting the
timescales available for condensation and causing deviations from
chemical equilibrium.
Regarding potential wind drivers, the need for high flux mean
opacities directs the search towards condensates with high radiative
cross sections in the wavelength range around the stellar flux
maximum, but also high abundances of the constituents to speed up
condensation, and arrive at sufficient total opacities. Furthermore,
by definition, a wind driver must be able to form close to the
stellar photosphere to accelerate the gas away from the star, which
requires grain materials with high condensation temperatures.

Looking at the most abundant condensible elements, the high bond
energy of the CO molecule -- blocking the less abundant of the two
elements -- leads to a convenient dichotomy in molecular and dust
chemistry for M- and C-type AGB stars. In the latter case, the
excess carbon will start forming amorphous carbon particles at high
temperatures, resulting in efficient radiative driving of an
outflow. For M-type stars with C/O $<$ 1, 
on the other hand, the excess oxygen needs to combine with elements
of lower abundances to form high-temperature condensates. Using the
criteria summarized above, silicates (olivine- and pyroxene-type
materials) suggest themselves as candidates, and seem to be
consistent with observed mid-IR spectral features. Specific points
are, however, still debated, e.g., if non-grey effects on grain
temperatures and condensation distances will suppress the inclusion
of Fe in silicate grains (cf. Woitke 2006b), if Mg-rich end members
of the olivine-/pyroxene-type material sequences provide sufficient
radiative pressure for driving an outflow (cf. H{\"o}fner 2008a), to
which degree the condensates are crystalline or amorphous, or what
particles are the seed nuclei for silicate condensation.

In contrast to the more or less satisfactory cases of C- and M-type
AGB stars, the enigmatic S stars still present a challenge to the
understanding of dust formation and outflows. Observations indicate
that these objects have wind characteristics which are similar to
the other two types (e.g. Ramstedt et al.~2006) but the lack of
abundant dust-forming elements in S-type giants (with both C and O
locked up in CO) makes it difficult to envision an efficient driving
mechanism, on par with, e.g., radiative pressure on amorphous carbon
grains. Various dust species have been suggested (cf. Ferrarotti \&
Gail 2002, 2006) but no detailed self-consistent dynamical models
are currently available.

The discussion of S-type AGB stars leads naturally into a topic that
could be called minor dust species, defined as the types of grains
that are no viable wind drivers due to low abundances or low
opacity. A considerable number of potential dust materials has been
inferred from the rich variety of features discovered in IR spectra
of cool giants but dynamical models are usually concentrating on a
handful of dominating species for computational reasons.
Nevertheless, some rare or low-opacity high-temperature condensates
may play a crucial role in the important process of grain nucleation
which is not well understood at present (see, e.g., Gail \& Sedlmayr
1999, Nuth \& Ferguson 2006, Patzer 2007, and references therein).

Having listed some present concerns about our understanding of dust
in evolved stars, the great progress achieved in recent years
through a combination of high-quality IR observations, laboratory
work on dust analogues, and detailed numerical modeling should not
be underestimated. A consistent, reasonably realistic description of
dust-driven mass loss of AGB stars seems within reach, both for C-
and M-type objects, providing urgently needed input for models of
stellar and galactic chemical evolution. Furthermore, the
exploration of intricate structures in circumstellar envelopes is
gaining momentum, both regarding interferometric observations and 3D
modeling, promising interesting new insights.

%%% THE BIBLIOGRAPHY
%%%
%%% CONSULT SECTION 3 OF "INSTRUCTIONS FOR AUTHORS" FOR HOW TO USE NATBIB.
%%% AUTHORS ARE ENCOURAGED TO USE EITHER THE "THEBIBLIOGRAPY" ENVIRONMENT
%%% BY UNCOMMENTING (DELETING THE "%" SYMBOL) THE COMMANDS BELOW, OR BY
%%% USING THE BIBTEX ENVIRONMENT. TO FIND OUT WHICH IS APPLICABLE TO YOUR
%%% CONTRIBUTION, CONSULT THE VOLUME EDITORS FOR YOUR PROCEEDINGS.
%%%

%\begin{thebibliography}{}
%\bibitem[]{}
%\bibitem[]{}
%\bibitem[]{}
%\bibitem[]{}
%\bibitem[]{}
%\bibitem[]{}
%\bibitem[]{}
%\bibitem[]{}
%\bibitem[]{}
%\bibitem[]{}
%\bibitem[]{}
%\bibitem[]{}
%\end{thebibliography}


\begin{thebibliography}{}

   \bibitem[2003]{andersen_etal03}
   Andersen A.C., H{\"o}fner  S., Gautschy-Loidl R., 2003, \aap, 400, 981

   \bibitem[]{}
   Begemann B., Dorschner J., Henning Th., et al. 1997, ApJ 476, 199

   \bibitem[]{}
   Boboltz D., Wittkowski M., 2005, ApJ 618, 953

   \bibitem[1983]{BH83}
   Bohren C.F., Huffman D., 1983, {\it Absorption and scattering of light
   by small particles}, John Wiley, New York

   \bibitem[]{b88}
   Bowen G.H. 1988, ApJ 329, 299

   \bibitem[]{dan94}
   Danchi W.C., Bester M., Degiacomi C.G., et al. 1994, AJ 107, 1469

   \bibitem[]{}
   Dominik C., Gail H.P., Sedlmayr E., Winters J.M. 1990, A\&A 240, 365

   \bibitem[]{dorschner03}
   Dorschner J. 2003, in: {\it Astromineralogy}, LNP 609,
   ed.~Th.~Henning, Spinger, 1

   \bibitem[]{}
   Fabian D., Posch T., Mutschke H., Kerschbaum F., Dorschner, J. 2001,
   A\&A 373, 1125

   \bibitem[]{}
   Ferrarotti A.S., Gail H.-P., 2001, A\&A 371, 133

   \bibitem[]{}
   Ferrarotti A.S., Gail H.-P., 2002, A\&A 382, 256

   \bibitem[]{}
   Ferrarotti A.S., Gail H.-P., 2006, A\&A 447, 553

   \bibitem[]{fgs91}
   Fleischer A.J., Gauger A., Sedlmayr E. 1991, A\&A 242, L1

   \bibitem[]{fgs92}
   Fleischer A.J., Gauger A., Sedlmayr E. 1992, A\&A 266, 321

   \bibitem[]{fgs95}
   Fleischer A.J., Gauger A., Sedlmayr E. 1995, A\&A 297, 543

   \bibitem[]{}
   Freytag B., H{\"o}fner S. 2008, A\&A 483, 571

   \bibitem[]{gail03}
   Gail H.-P. 2003, in: {\it Astromineralogy}, LNP 609,
   ed.~Th.~Henning, Spinger, p.55

   \bibitem[]{gs88}
   Gail H.-P., Sedlmayr E. 1988, A\&A 206, 153

   \bibitem[1999]{GS99}
   Gail H.-P., Sedlmayr E. 1999, A\&A 347, 594

   \bibitem[]{ggs90}
   Gauger A., Gail H.-P., Sedlmayr E. 1990, A\&A 235, 345

   \bibitem[2004]{loidl_etal04}
   Gautschy-Loidl R., H{\"o}fner S., J{\o}rgensen U.G., Hron J., 2004, \aap, 422, 289

   \bibitem{GDetal03}
   Gonz{\'a}lez Delgado D., Olofsson H., Kerschbaum F., et al.,
     % Sch{\"o}ier, F.~L., Lindqvist, M., Groenewegen, M.~A.~T.,
   2003, A{\&}A 411, 123

   \bibitem[2004]{gh04}
   Gustafsson B., H{\"o}fner S. 2004,
   in: {\it Asymptotic Giant Branch Stars}, Habing H.J., Olofsson H. (eds.),
   Springer, p.149

   \bibitem[2004]{ho04}
   Habing H.J., Olofsson H. 2004,
   in: {\it Asymptotic Giant Branch Stars}, Habing H.J., Olofsson H. (eds.),
   Springer, p.1

   \bibitem[]{hbsw00}
   Hofmann K.-H., Balega Y., Scholz M., Weigelt G. 2000, A\&A 353, 1016

   \bibitem[2007]{hoefner07}
   H{\"o}fner S., 2007, ASP Conf.~Ser.~378, p.~145

   \bibitem[2008a]{hoefner08a}
   H{\"o}fner S., 2008a, \aap, 491, L1

   \bibitem[2008b]{hoefner08b}
   H{\"o}fner S., 2008b, Phys.~Scr., T133

   \bibitem[]{hd97}
   H{\"o}fner S., Dorfi E.A. 1997, A{\&}A 319, 648

   \bibitem[]{hfd95}
   H{\"o}fner S., Feuchtinger M., Dorfi E.A. 1995, A\&A 297, 815

   \bibitem[2003]{hoefner_etal03}
   H{\"o}fner S., Gautschy-Loidl R., Aringer B., J{\o}rgensen U.G., 2003, \aap, 399, 589

   \bibitem[2003]{jager_etal03}
   J{\"a}ger C., Dorschner J., Mutschke H., Posch Th., Henning Th., 2003, \aap, 408, 193

   \bibitem{jager_etal98}
   J{\"a}ger C., Mutschke H., Henning Th., 1998, A{\&}A, 332, 291

   \bibitem[2003]{jwls03}
   Jeong K. S., Winters J. M., Le Bertre T., Sedlmayr E. 2003, A\&A, 407, 191

   \bibitem[2003]{Kruegel03}
   Kr{\"u}gel E., 2003, {\it The Physics of Interstellar Dust}, IoP

   \bibitem[2004]{lw04}
   Lattanzio J.C., Wood P.R. 2004,
   in: {\it Asymptotic Giant Branch Stars}, Habing H.J., Olofsson H. (eds.),
   Springer, p.23

   \bibitem[]{}
   Le Bertre T., Winters J.M. 1998, A\&A 334, 173

   \bibitem[]{}
   Mattsson L., H{\"o}fner S., Herwig F., 2007, A\&A 470, 339

   \bibitem{2008A&A...484L...5M}
   Mattsson L., Wahlin R., H{\"o}fner S., Eriksson K., 2008, A\&A, 484, L5

   \bibitem[]{mh99}
   Mauron N., Huggins P.J. 1999, A\&A 349, 203

   \bibitem[]{mh00}
   Mauron N., Huggins P.J. 2000, A\&A 359, 707

   \bibitem[]{mw03}
   Molster F.J., Waters L.B.F.M. 2003, in: {\it Astromineralogy}, LNP 609,
   ed.~Th.~Henning, Spinger, p.121

   \bibitem[]{mol02a}
   Molster F.J., Waters L.B.F.M., Tielens A.G.G.M, 2002a,
   A\&A 382, 222

   \bibitem[]{mol02b}
   Molster F.J., Waters L.B.F.M., Tielens A.G.G.M, Barlow M.J. 2002b,
   A\&A 382, 184

   \bibitem[]{mol02c}
   Molster F.J., Waters L.B.F.M., Tielens A.G.G.M, et al. %Koike C., Chihara H.
   2002c,
   A\&A 382, 241

   \bibitem[]{mon04}
   Monnier J.D., Millan-Gabet R., Tuthill P.G., et al. 2004, ApJ 605, 436

   \bibitem[]{CO_1}
   Nowotny W., Aringer B., H{\"o}fner S., et al. %Gautschy-Loidl R., Windsteig W.
   2005a, A\&A, 437, 273

   \bibitem[]{CO_2}
   Nowotny W., Lebzelter T., Hron J., H{\"o}fner S. 2005b, A\&A, 437, 285

   \bibitem[]{}
   Nuth J.A., Ferguson F.T. 2006, ApJ 649, 1178

   \bibitem{Olofsson02}
   Olofsson H., Gonz{\'a}lez Delgado D., Kerschbaum F., Sch{\"o}ier F.~L., 2002, A{\&}A 391, 1053

   \bibitem[]{}
   Onaka T., de Jong T., Willems F.J. 1989, A\&A 218, 169

   \bibitem[2007]{patzer07}
   Patzer A.B.C., 2007, ASP Conf.~Ser.~378, p.~181

   \bibitem[]{}
   Posch T., Kerschbaum F., Mutschke H., et al. % Fabian D., Dorschner J., Hron J.
   1999,
   A\&A 352, 609

   \bibitem []{}
   Ragland S., Le Coroller H., Pluzhnik E., et al., 2008, ApJ 679, 746

   \bibitem[2006]{ramstedt}
   Ramstedt S., Sch{\"o}ier F. L., Olofsson H., Lundgren A. A., 2006, A\&A, 454, L103

   \bibitem[]{}
   Rouleau F., Martin P.G. 1991, ApJ 377, 526

   \bibitem[]{cssh3}
   Sandin C., H{\"o}fner S. 2004, A\&A 413, 789

   \bibitem[]{kps03}
   Schr{\"o}der K.-P., Wachter A., Winters J. M. 2003, A\&A 398, 229

   \bibitem[]{kps99}
   Schr{\"o}der K.-P., Winters J. M., Sedlmayr E. 1999, A\&A 349, 898

   \bibitem[]{sid01}
   Simis Y.J.W., Icke V., Dominik C. 2001, A\&A 371, 205

   \bibitem[]{}
   Tatebe K., Wishnow E. H., Ryan C. S., et al. 2008, ApJ 689, 1289

   \bibitem[]{tlsw03a}  
   Tej A., Lan\c{c}on A., Scholz M. 2003, A\&A 401, 347

   \bibitem[]{}
   Tevousjan S., Abdeli K.-S., Weiner J. et al. 2004, 611, 466

   \bibitem[]{}
   Vardya M.S., De Jong T., Willems F.J. 1986, ApJ 304, L29

   \bibitem[]{wach02}
   Wachter A., Schr{\"o}der K.-P., Winters, J. M., et al.
%  Arndt, T. U., Sedlmayr, E.
   2002, A\&A 384, 452

   \bibitem[]{wach08}
   Wachter A.,  Winters, J. M., Schr{\"o}der K.-P., Sedlmayr, E.
   2008, A\&A 497, 504

   \bibitem[]{weig98}
   Weigelt G., Balega Y.Y., Bl{\"o}cker T., et al. 1998, A\&A 333, L51

   \bibitem[]{weig02}
   Weigelt G., Balega Y.Y., Bl{\"o}cker T., et al. 2002, A\&A 392, 131

   \bibitem[]{wein04}
   Weiner J. 2004, ApJ 611, L37

   \bibitem[]{wfgs}
   Winters J.M., Fleischer A.J. Gauger A., Sedlmayr E. 1994,
   A{\&}A 290, 623

   \bibitem[]{wfls97}
   Winters J.M., Fleischer A.J. Le Bertre T., Sedlmayr E. 1997,
   A{\&}A 326, 305

   \bibitem[]{wljhs00}
   Winters J.M., Le Bertre T., Jeong, K.S., Helling C., Sedlmayr E. 2000,
   A{\&}A 361, 641

   \bibitem[]{}
   Wittkowski M., Boboltz D. A., Driebe T., et al. 2008, A\&A 479, L21

   \bibitem[]{}
   Wittkowski M., Boboltz D. A., Ohnaka K., et al. 2007, A\&A 470, 191

   \bibitem[]{}
   Woitke P. 2006a, A\&A, 452, 537

   \bibitem[2006]{woitke}
   Woitke P. 2006b, A\&A, 460, L9

   \bibitem[]{w79}
   Wood P.R. 1979, ApJ 227, 220

   \bibitem[]{}
   Woodruff H.C., Tuthill P.G., Monnier J.D., et al. 2008, ApJ 673, 418

\end{thebibliography}
\end{document}